\documentclass{elsart}
\usepackage{epsfig}
\newlength{\wordlength}

\newlength{\onewordlength}

    \newcommand{\ba}{\begin{eqnarray}}
    \newcommand{\ea}{\end{eqnarray}}
    \newcommand{\be}{\begin{equation}}
    \newcommand{\ee}{\end{equation}}

    \newcommand{\AmS}{{\protect\the\textfont2%
  A\kern-.1667em\lower.5ex\hbox{M}\kern-.125emS}}

\newcommand{\bn}{{\bf n}}

\newcommand {\bk} {{\mathbf k}}
\newcommand {\bfr} {{\mathbf r}}

\newcommand{\calH}{{\mathcal H}}

\newcommand{\calM}{{\mathcal M}}
\newcommand{\calZ}{{\mathcal Z}}
\newcommand{\calY}{{\mathcal Y}}

\newtheorem{theorem}{Theorem}

\begin{document}
\runauthor{PKU}
\begin{frontmatter}

\title{Two Particle States and the $S$-matrix Elements in Multi-channel Scattering}
\author[PKU]{Song He},
\author[PKU]{Xu Feng}
\author[PKU]{and Chuan Liu}

\address[PKU]{School of Physics\\
          Peking University\\
                  Beijing, 100871, P.~R.~China}
%\address[ITP]{Institute of Theoretical Physics\\
%                Academia Sinica\\
%                Beijing, 100080, P.~R.~China}
%\address[IHEP]{Institute of High Energy Physics\\
%                Academia Sinica\\
%                P.~O.~Box 918\\
%                Beijing, 100039, P.~R.~China}
                \thanks{This work is supported by the National Natural
 Science Foundation (NFS) of China under grant
 No. 10421003, No. 10235040 and supported by the
 Trans-century fund from Chinese
 Ministry of Education.}

 \begin{abstract}
 Using a quantum mechanical model,
 the exact energy eigenstates for two-particle two-channel
 scattering are studied in a cubic box
 with periodic boundary conditions in all three
 directions. A relation between the exact energy eigenvalue in
 the box and the two-channel $S$-matrix elements in the continuum
 is obtained. This result can be viewed as a
 generalization of the well-known L\"uscher's formula which
 establishes a similar relation in elastic scattering.
 \end{abstract}
 \end{frontmatter}

%\newpage

 \section{Introduction}

 Scattering experiments play an important role
 in the study of interactions among particles.
 In these experiments, scattering cross sections are
 measured. By a partial wave analysis, one obtains the
 experimental results on particle-particle scattering
 in terms scattering phase shifts in channels of
 definite quantum numbers.
 In the case of strong interaction, experimental results on
 hadron-hadron scattering phase shifts are available in the
 literature~\cite{E86500:pipi,matison74:Kpi_exp_a,johannesson73:Kpi_exp_b,%
 shaw80:Kpi_exp_c,martin81:KN}.
 On the theoretical side, Quantum Chromodynamics (QCD)
 is believed to be the underlying theory of strong interactions. However, due
 to its non-perturbative nature, low-energy hadron-hadron
 scattering should be studied with a non-perturbative
 method. Lattice QCD provides a genuine non-perturbative method which
 can tackle these problems in principle, using numerical simulations.
 In a typical lattice calculation, energy eigenvalues of two-particle states
 with definite symmetry can be obtained by measuring appropriate
 correlation functions. Therefore, it would be desirable to
 relate these energy eigenvalues which are available through
 lattice calculations to the scattering phases which are obtained
 in the scattering experiment.  This was accomplished
 in a series of papers by L\"uscher
 \cite{luscher86:finiteb,luscher90:finite,luscher91:finitea,luscher91:finiteb}
 for a cubic box topology.
 In these references, especially Ref.~\cite{luscher91:finitea},
 L\"uscher found a non-perturbative relation of
 the energy of a two-particle state in a cubic box (a
 torus) with the corresponding elastic scattering phases of the two
 particles in the continuum. This formula, now known as
 L\"uscher's formula, has been utilized in a number of
 applications, e.g. linear sigma model in the broken phase \cite{Zimmermann94},
 and also in quenched QCD~\cite{gupta93:scat,fukugita95:scat,jlqcd99:scat,%
 JLQCD02:pipi_length,chuan02:pipiI2,juge03:pipi_length,CPPACS03:pipi_phase,ishizuka03:pipi_length}.
 Due to limited numerical computational power,
 the $s$-wave scattering length, which is related to the scattering
 phase shift at vanishing relative three momentum,
 is mostly studied in hadron scattering using quenched approximation.
 CP-PACS collaboration calculated the scattering
 phases at non-zero momenta in pion-pion $s$-wave scattering in the $I=2$
 channel \cite{CPPACS03:pipi_phase} using quenched Wilson
 fermions and recently also in two flavor
 full QCD \cite{CPPACS03:pipi_phase_unquench}.

 For hadron scattering at low energies, usually the
 elastic scattering is dominant since the inelastic channels
 are not opened. However, when the energy of the scattering
 process exceeds some threshold, inelastic scattering starts
 to contribute and the scattering of the particles cannot
 be described by single channel elastic scattering anymore.
 In the case of pion-pion scattering, for example, the scattering
 process is elastic below the four pion and the two kaon
 threshold. If the center of mass energy exceeds the four pion
 threshold, inelastic effects starts to contribute. The inelastic
 effects become very important when the energy is getting close
 to the two kaon threshold. At this point, pion-pion can be
 scattered into kaon-kaon pair final state.
 Although the four pion threshold is in fact below the
 two kaon threshold, four pion final state will not
 contribute significantly due to its weak coupling to
 the two pion initial state. The fact that four pion states
 are coupled to the two pion state weakly in the low-energy limit
 is seen from the QCD chiral lagrangian. A six pion vertex in this
 lagrangian involves derivative coupling which is vanishing
 in the low-energy limit.
 Experimental investigations show that the contribution of
 four pion states would only make substantial contribution
 when the energy is well over 1GeV.
 Therefore, if the energy is not much higher than 1GeV,
 pion-pion scattering can be approximated rather well by a
 two-channel model.
 It is then interesting to study the relation between the
 multi-channel two particle states
 and the scattering phases, just as what we have done in
 the single channel case.

 In this paper, we establish a relation between the energy
 of a two particle state in a finite cubic box and the scattering
 matrix parameters. It is a generalization of the famous L\"uscher formula
 to the multi-channel situation.
 This relation is non-perturbative in nature and
 it is derived in a quantum mechanical model
 of two-channel scattering. The result can also be
 generalized to the case of asymmetric box.
 Further generalization to the
 case of massive field theory is under consideration.

 \section{The quantum mechanical model to two channel scattering}
 \label{sec:model}

 In this paper, we study a quantum mechanical model of
 two-channel scattering. Generalization to more channels
 can be done similarly. The model under investigation
 has the following Hamiltonian:
 \be
 \label{eq:hamiltonian}
 H=\left(\begin{array}{cc}
 -{1\over 2m_1}\nabla^2 & 0 \\
 0 & E_T-{1\over 2m_2}\nabla^2\end{array}\right)
 +\left(\begin{array}{cc}
 V_1(r)      & \Delta(r) \\
 \Delta^*(r) & V_2(r)\end{array}\right)\;.
 \ee
 We will also use the notation:
 \be
 \label{eq:matrix_potential}
 V(r)=\left(\begin{array}{cc}
 V_1(r)      & \Delta(r) \\
 \Delta^*(r) & V_2(r)\end{array}\right)\;
 \ee
 to denote the matrix valued potential
 whose matrix elements $V_1(r)$, $V_2(r)$
 and $\Delta(r)$ all vanish for $r>R$.
 $E_T>0$ designates a positive threshold energy
 of the second channel. That is to say, if the center of
 mass energy for the scatter process is less than $E_T$,
 there will be no asymptotic scattering states in the
 second channel.

 Energy eigenstates of the Hamiltonian~(\ref{eq:hamiltonian}) with
 energy $E$ can be decomposed into spherical harmonics:
 \be
 \Psi(\bfr)=\sum_{l,m} Y_{lm}(\hat{\bfr})
 \left(\begin{array}{c}
 \psi_{1;lm}(r)\\
 \psi_{2;lm}(r)\end{array}\right)
 \;.
 \ee
 The radial wave-functions $\psi_{i;lm}(r)$ with $i=1,2$ satisfy
 the radial Schr\"odinger equation:
 \ba
 \label{eq:radial_schrodinger}
 \left[{1\over 2m_1}\left(
 {d^2\over dr^2}+{2\over r}{d\over dr}-{l(l+1)\over r^2}\right)
 +E-V_1(r)\right]\psi_1(r)&=&\Delta(r)\psi_2(r) \;,\nonumber \\
 \left[{1\over 2m_2}\left(
 {d^2\over dr^2}+{2\over r}{d\over dr}-{l(l+1)\over r^2}\right)
 +E'-V_2(r)\right]\psi_2(r)&=&\Delta^*(r)\psi_1(r) \;,
 \ea
 where $E'\equiv E-E_T$.

 About the coupled differential
 equations~(\ref{eq:radial_schrodinger}),
 the following statement can be proven.
 \begin{theorem}
 If the matrix valued potential~(\ref{eq:matrix_potential}) is
 such that every matrix element of $r^2V(r)$ is
 analytic around $r=0$ and that
 $\lim_{r\rightarrow 0}r^2V(r)=0$, then the
 coupled differential equations~(\ref{eq:radial_schrodinger})
 has two finite, linearly independent solutions near $r=0$:
 $u^{(i)}_{l;j}(r)$, with $i=1,2$ designating different
 solutions and $j=1,2$ denoting different component of the solution.
 Moreover, these solutions can be chosen such that:
 \be
 \label{eq:u_def}
 u^{(i)}_{l;j}(r) \sim r^l\delta_{ij}\;.
 \ee
  Equations~(\ref{eq:radial_schrodinger}) has two
 further linear independent solutions $v^{(i)}_{l;j}(r)$
 that are unbounded near $r=0$ and satisfy:
 $v^{(i)}_{l;j}(r) \sim r^{-l-1}\delta_{ij}$.
 \end{theorem}
 The proof of this theorem is quite similar to the
 proof in the single channel case. Details are provided
 in the appendix.

 \section{Two channel scattering and the $S$ matrix in infinite volume}
 \label{sec:2channel}

 In this section, we briefly describe the quantum mechanical
 treatment of inelastic scattering. We will concentrate on
 the two channel case, although the formalism can easily
 be generalized to more channels.

 At large $r$ where the potential $V(r)$ vanishes, the
 wave function of the scattering state can be chosen to
 have the following form:
 \be
 \label{eq:psi1}
 \Psi^{(1)}(\bfr)\stackrel{r\rightarrow\infty}{\longrightarrow}
 \left(\begin{array}{c}
 e^{i\bk_1\cdot\bfr}+
 f_{11}(\hat{\bk}_1\cdot\hat{\bfr}){e^{ik_1r}\over r}\\
 f_{21}(\hat{\bk}_1\cdot\hat{\bfr})\sqrt{{m_2\over m_1}}
 {e^{ik_2r}\over r}\end{array}\right)\;.
 \ee
 This wave function has the property that in the
 remote past, it becomes an incident plane wave
 in the first channel with definite wave vector $\bk_1$.
 It is an eigenstate
 of the full Hamiltonian with energy: $E=\bk^2_1/(2m_1)$.
 Similarly, if the energy $E>E_T$, one can also
 build an eigenstate of the Hamiltonian which in the
 remote pase becomes an incident plane wave in the
 second channel:
 \be
 \label{eq:psi2}
 \Psi^{(2)}(\bfr)\stackrel{r\rightarrow\infty}{\longrightarrow}
 \left(\begin{array}{c}
 f_{12}(\hat{\bk}_2\cdot\hat{\bfr})\sqrt{{m_1\over m_2}}
 {e^{ik_1r}\over r}\\
 e^{i\bk_2\cdot\bfr}+
 f_{22}(\hat{\bk}_2\cdot\hat{\bfr}){e^{ik_2r}\over r}
 \end{array}\right)\;.
 \ee
 This state is also an eigenstate of the Hamiltonian
 with energy $E=E_T+\bk^2_2/(2m_2)$.

 In partial wave analysis, one decompose the
 coefficients: $f_{ij}$ appearing in the above
 construction into spherical harmonics:
 \ba
 f_{11}(\hat{\bk}_1\cdot\hat{\bfr}) &=&
 {1\over 2ik_1}\sum^\infty_{l=0}(2l+1)
 \left[S^{(l)}_{11}(k_1)-1\right]
 P_l(\hat{\bk}_1\cdot\hat{\bfr}) \;,\\
 f_{21}(\hat{\bk}_1\cdot\hat{\bfr}) &=&
 {1\over 2i\sqrt{k_1k_2}}\sum^\infty_{l=0}(2l+1)
 S^{(l)}_{21}(k_1)
 P_l(\hat{\bk}_1\cdot\hat{\bfr}) \;,
 \ea
 and similar expressions for $f_{12}$ and $f_{22}$.
 With this we find the two scattering eigenstates
 in Eq.~(\ref{eq:psi1}) and Eq.~(\ref{eq:psi2})
 can be expressed as:
 \ba
 \label{eq:Ypsi}
 \Psi^{(1)}(\bfr)&\stackrel{r\rightarrow\infty}{\longrightarrow}&
 \sum_{lm} 4\pi Y_{lm}(\hat{\bfr})Y^*_{lm}(\hat{\bk}_1)
 \left(\begin{array}{c}
 {1\over 2ik_1r}\left[
 S^{(l)}_{11}e^{ik_1r}+(-)^{l+1}e^{-ik_1r}\right] \\
 {1\over 2i\sqrt{k_1k_2}r}\sqrt{{m_2\over m_1}}
 S^{(l)}_{21}e^{ik_2r}
 \end{array}\right)\;, \nonumber \\
 \Psi^{(2)}(\bfr)&\stackrel{r\rightarrow\infty}{\longrightarrow}&
 \sum_{lm} 4\pi Y_{lm}(\hat{\bfr})Y^*_{lm}(\hat{\bk}_2)
 \left(\begin{array}{c}
 {1\over 2i\sqrt{k_1k_2}r}\sqrt{{m_1\over m_2}}
 S^{(l)}_{12}e^{ik_1r}\\
 {1\over 2ik_2r}\left[
 S^{(l)}_{22}e^{ik_2r}+(-)^{l+1}e^{-ik_2r}\right]
 \end{array}\right)\;.
 \ea
 The two component wave-functions appearing in the
 above formulae are in fact the radial wave functions
 of the Schr\"odinger equation~(\ref{eq:radial_schrodinger})
 in the large $r$ region:
 \footnote{In fact, if we use the corresponding spherical
 Bessel's function $j_l$ and $n_l$, we can get an expression
 for $w^{(1)}_l(r)$ and $w^{(2)}_l(r)$ for $r>R$.
 Eq.~(\ref{eq:w_def}) is the asymptotic form when
 $r\rightarrow\infty$.}
 \ba
 \label{eq:w_def}
 w^{(1)}_l(r) &\simeq& \left(\begin{array}{c}
 {1\over 2ik_1r}\left[
 S^{(l)}_{11}e^{ik_1r}+(-)^{l+1}e^{-ik_1r}\right] \\
 {1\over 2i\sqrt{k_1k_2}r}\sqrt{{m_2\over m_1}}
 S^{(l)}_{21}e^{ik_2r}
 \end{array}\right)\;, \nonumber \\
 w^{(2)}_l(r) &\simeq& \left(\begin{array}{c}
 {1\over 2i\sqrt{k_1k_2}r}\sqrt{{m_1\over m_2}}
 S^{(l)}_{12}e^{ik_1r}\\
 {1\over 2ik_2r}\left[
 S^{(l)}_{22}e^{ik_2r}+(-)^{l+1}e^{-ik_2r}\right]
 \end{array}\right)\;.
 \ea
 It is obvious that the two radial wave functions
 $w^{(1)}_l(r)$ and $w^{(2)}_l(r)$ are
 linearly independent. Therefore, according to the theorem
 stated in the previous section, they are linear superpositions of the
 general solutions: $u^{(1)}_l(r)$ and $u^{(2)}_l(r)$.
 The converse is also true. The radial wave functions
 $u^{(1)}_l(r)$ and $u^{(2)}_l(r)$ defined
 via Eq.~(\ref{eq:u_def}) can be expressed as
 linear superpositions of the two
 radial wave functions in Eq.~(\ref{eq:w_def}).
 In other words, there exists a non-singular
 $2\times 2$ matrix $C$ such that:
 \be
 w^{(i)}_l(r)=\sum_j C_{ij}u^{(j)}_l(r) \;,\;\;
 u^{(i)}_l(r)=\sum_j C^{-1}_{ij}w^{(j)}_l(r)\;.
 \ee

 Another important physical property of the
 wave-function~(\ref{eq:Ypsi}) is that the
 matrix elements which enter the expansion, namely
 $S^{(l)}_{ij}$, form a $2\times 2$ unitary matrix which
 is nothing but the $S$-matrix in the subspace with
 orbital angular momentum $l$.
 This unitarity condition comes directly from
 the probability conservation law of quantum mechanics.
 We have, for example:
 \be
 |S^{(l)}_{11}|^2+|S^{(l)}_{21}|^2=1\;,
 \;\;
 |S^{(l)}_{12}|^2+|S^{(l)}_{22}|^2=1\;.
 \ee
 In practice, if the theory has $CP$ symmetry which is
 the case in QCD, the two-channel $S$-matrix is usually
 parameterized as:
 \footnote{For the potential models, we assume that the
 potential is invariant under time reversal and parity. Then,
 one has $S_{fi}=S_{i^*f^*}$, where $i^*$ and $f^*$ denotes
 the time-reversed state of $i$ and $f$. With this
 in mind, it is easily seen that
 eq.~(\ref{eq:S_parametrize}) is the most general $2\times 2$
 unitary matrix for the scattering matrix of spinless particles.}
 \be
 \label{eq:S_parametrize}
 S^{(l)}(E)=\left(\begin{array}{cc}
 \eta_l e^{2i\delta^l_1} &
 i\sqrt{1-\eta^2_l}e^{i(\delta^l_1+\delta^l_2)} \\
 i\sqrt{1-\eta^2_l}e^{i(\delta^l_1+\delta^l_2)} &
 \eta_l e^{2i\delta^l_2} \end{array}
 \right)\;,
 \ee
 where the real parameters: $\delta^l_1(E)$, $\delta^l_2(E)$
 and $\eta_l(E)$ are all functions of the energy $E$.
 We will assume in the following that the $S$-matrix of
 the scattering problem has this form.

 \section{Energy eigenfunctions on a torus}
 \label{sec:2channel_finiteV}

 Now we enclose the system discussed above in a cubic, periodic
 box with finite extension $L$ in every spatial direction.
 The Schr\"odinger equation for the system now takes a
 similar form as in the infinite volume except that the
 potential is periodically extended and the
 (two-component) eigenfunction has to satisfy the
 periodic boundary condition:
 \be
 \label{eq:schrodinger_finite}
 [H_0+V_L(\bfr)]\psi(\bfr)=E\psi(\bfr)\;,
 \;\; \psi(\bfr+L\bn)=\psi(\bfr)\;.
 \ee
 where the free Hamiltonian is given by:
 \be
 H_0=\left(\begin{array}{cc}
 -{1\over 2m_1}\nabla^2 & 0 \\
 0 & E_T-{1\over 2m_2}\nabla^2\end{array}\right)
 \ee
 and the periodically extended potential is:
 \be
 V_L(\bfr)\equiv \sum_\bn V(\bfr+L\bn)\;.
 \ee
 The eigenvalue equation~(\ref{eq:schrodinger_finite}) now
 has discrete spectrum and the corresponding
 eigenfunctions are smooth.

 It is convenient to partition the whole space into two
 regions. In the inner region, every point satisfies the
 condition: $|\bfr|<R$, $mod(L)$. In the outer region:
 \be
 \Omega=\{\bfr|: |\bfr|>R\;,mod(L)\}\;.
 \ee
 Note that in the outer region $\Omega$,
 the interaction potential $V_L(\bfr)=0$
 and the Schr\"odinger equation~(\ref{eq:schrodinger_finite})
 reduces to two {\em decoupled} Helmholtz equations:
 \be
 (\nabla^2+k^2_i)\psi_i(\bfr)=0\;,
 \;\; \mbox{i=1,2}\;,
 \ee
 where the energy eigenvalue $E$ is given by:
 \be
 \label{eq:energy}
 E={k^2_1\over 2m_1}=E_T+{k^2_2\over 2m_2}\;.
 \ee
 For energy $0<E<E_T$, $k_1$ is real and $k_2$ takes
 purely imaginary values;
 for energy $E<0$, both $k_1$ and $k_2$ are purely
 imaginary.; for $E>E_T$, which is the case of
 two channel scattering above the threshold, both
 $k_1$ and $k_2$ are real numbers.

 It it easy to see that:
 \be
 \Psi(\bfr;E)=\sum_{lm}
 \left[\sum^2_{i=1} b^{(i)}_{lm}u^{(i)}_l(r)\right]
  Y_{lm}(\bn)\;.
 \ee
 solves the Schr\"odinger equation in the
 inner region for $|\bfr|<R$.
 In the outer region $\Omega$, for a given value of
 energy $E$, the corresponding eigenfunction must be
 superposition of the free Schr\"odinger equation,
 which in the outer region decouples to two
 independent Helmholtz equations.
 Since there are two linear independent radial wave
 functions, the eigenfunction must be some linear
 combination of the two:
 \be
 \label{eq:outer_w}
 \Psi(\bfr;E)=\sum_{lm}
 \left[\sum_{i}c^{(i)}_{lm}w^{(i)}_l(r)\right]
 Y_{lm}(\bn)\;, \mbox{ for $r > R$.}\;,
 \ee
 with $c^{(j)}_{lm}=\sum_{i}b^{(i)}_{lm}C_{ij}$ being
 non-vanishing coefficients.
 Note that when the system is enclosed in a finite periodic
 box, the exact energy eigenvalues become discrete.
 The degeneracy in the radial wave-function in general
 is then lifted. That is to say, for a given value of energy,
 there exists only one radial wave-function, unlike in
 the infinite volume where there are two such
 wave-functions for a given energy.

 On the other hand, in the region $\Omega$, the solution
 must be linear superposition of the singular periodic
 solutions of Helmholtz equation:
 \be
 \label{eq:outer_expand}
 \Psi(\bfr;E)=\left(\begin{array}{c}
 \sum_{lm} v^{(1)}_{lm}G_{lm}(\bfr;k^2_1) \\
 \sum_{lm} v^{(2)}_{lm}G_{lm}(\bfr;k^2_2)
 \end{array}\right)\;.
 \ee
 Combining Eq.~(\ref{eq:outer_expand}) and
 Eq.~(\ref{eq:outer_w}), using the
 basic expansion of $G_{lm}(\bfr)$:
 \be
 \label{eq:expand_Ylm}
 G_{lm}(\bfr;k^2)={(-)^lk^{l+1}\over 4\pi}
 \left[Y_{lm}(\Omega_\bfr)n_l(kr)
 +\sum_{l'm'}\calM_{lm;l'm'}Y_{l'm'}(\Omega_\bfr)
 j_{l'}(kr)\right]\;,
 \ee
 we arrive at the following set of linear
 equations:
 \ba
 \label{eq:big_linear}
  c^{(1)}_{lm}(S^{(l)}_{11}+1)
 +c^{(2)}_{lm}\sqrt{{k_1m_1\over k_2m_2}}S^{(l)}_{12}
 &=& \sum_{l'm'}{(-)^{l'}k^{l'+1}_1\over 4\pi}
 v^{(1)}_{l'm'}\calM^{(1)}_{l'm';lm} \;,
 \nonumber \\
 -ic^{(1)}_{lm}(S^{(l)}_{11}-1)
 -ic^{(2)}_{lm}\sqrt{{k_1m_1\over k_2m_2}}S^{(l)}_{12}
 &=& {(-)^{l}k^{l+1}_1\over 4\pi}
 v^{(1)}_{lm}\;, \nonumber \\
  c^{(2)}_{lm}(S^{(l)}_{22}+1)
 +c^{(1)}_{lm}\sqrt{{k_2m_2\over k_1m_1}}S^{(l)}_{21}
 &=& \sum_{l'm'}{(-)^{l'}k^{l'+1}_2\over 4\pi}
 v^{(2)}_{l'm'}\calM^{(2)}_{l'm';lm} \;,
 \nonumber \\
 -ic^{(2)}_{lm}(S^{(l)}_{22}-1)
 -ic^{(1)}_{lm}\sqrt{{k_2m_2\over k_1m_1}}S^{(l)}_{21}
 &=& {(-)^{l}k^{l+1}_2\over 4\pi}
 v^{(2)}_{lm}\;.
 \ea
 In these equations, the symbol $\calM^{(i)}_{l'm';lm}$
 represents $\calM_{l'm';lm}(k^2_i)$ for $i=1,2$,
 respectively. The explicit expression for
 for $\calM_{l'm';lm}(k^2_i)$ are given in
 Ref.~\cite{luscher91:finitea} which we quote here:
 \ba
 \label{eq:calM-zeta}
 \calM_{lm;js}(k^2) &=&
 \sum_{l'm'}{(-)^si^{j-l}\calZ_{l'm'}(1,q^2)\over
 \pi^{3/2}q^{l'+1}}
 \sqrt{(2l+1)(2l'+1)(2j+1)}
 \nonumber \\
 &\times &
 \left(\begin{array}{ccc}
 l & l' & j \\
 0 & 0  & 0 \end{array}
 \right)
 \left(\begin{array}{ccc}
 l & l' & j \\
 m & m' & -s \end{array}
 \right)\;.
 \ea
 Here we have used the Wigner's $3j$-symbols and
 $q=kL/(2\pi)$. The zeta function $\calZ_{lm}(s,q^2)$ is
 defined as:
 \be
 \label{eq:zeta_def}
 \calZ_{lm}(s,q^2)=
 \sum_{\bn} {\calY_{lm}(\bn) \over (\bn^2-q^2)^s}\;.
 \ee
 According to this definition,
 the summation at the right-hand side of Eq.~(\ref{eq:zeta_def})
 is formally divergent for $s=1$ and needs to be analytically continued.
 Following similar discussions as in Ref.~\cite{luscher91:finitea},
 one could obtain a finite expression for the zeta
 function which is suitable for
 numerical evaluation~\cite{chuan04:asymmetric,chuan04:asymmetric_long}.
 From the analytically continued formula,
 it is obvious from the symmetry of $O(Z)$
 that, for $l\leq 4$, the only non-vanishing zeta functions
 at $s=1$ are: $\calZ_{00}$, and $\calZ_{40}$.

 Eliminating the coefficients $v^{(1)}_{lm}$ and $v^{(2)}_{lm}$
 from the set of equations~(\ref{eq:big_linear}) , one
 obtains a homogeneous linear equation for the
 coefficients $c^{(1)}_{lm}$ and $c^{(2)}_{lm}$.
 In order to have non-trivial solutions for
 the coefficients $c^{(1)}_{lm}$ and $c^{(2)}_{lm}$,
 the corresponding matrix has to be singular.
 This condition then gives:
 \be
 \label{eq:main_result}
 \left|\begin{array}{cc}
 \calM^{(1)+}_{l'm';lm}-
 S^{(l')}_{11}\calM^{(1)-}_{l'm';lm}
 &
 \sqrt{{k_2m_2\over k_1m_1}}S^{(l')}_{21}
 \calM^{(2)-}_{l'm';lm}
 \\
 \sqrt{{k_1m_1\over k_2m_2}}S^{(l')}_{12}
 \calM^{(1)-}_{l'm';lm} &
 \calM^{(2)+}_{l'm';lm}-
 S^{(l')}_{22}\calM^{(2)-}_{l'm';lm}
 \end{array}\right|=0\;,
 \ee
 where the matrix elements $\calM^{(i)\pm}_{l'm';lm}$ are
 defined as:
 \be
 \calM^{(i)\pm}_{l'm';lm}(k^2_i)
 =\calM^{(i)}_{l'm';lm}(k^2_i)
 \pm i\delta_{l'l}\delta_{m'm}\;,
 \ee
 with the parameter $k^2_i$ related to
 the exact energy eigenvalue via Eq.~(\ref{eq:energy}).
 Now if we further assume that the matrices:
 $\calM^{(i)-}_{l'm';lm}(k^2_i)$ are non-singular,
 Eq.~(\ref{eq:main_result}) may be expressed as:
 \be
 \label{eq:main_resultU}
 \left|\begin{array}{cc}
 U^{(1)}_{l'm';lm}-
 S^{(l)}_{11}\delta_{l'l}\delta_{m'm}
 &
 \sqrt{{k_2m_2\over k_1m_1}}S^{(l)}_{21}
 \delta_{l'l}\delta_{m'm}
 \\
 \sqrt{{k_1m_1\over k_2m_2}}S^{(l)}_{12}
 \delta_{l'l}\delta_{m'm}
 &
 U^{(2)}_{l'm';lm}- S^{(l)}_{22}\delta_{l'l}\delta_{m'm}
 \end{array}\right|=0\;,
 \ee
 where the unitary matrices $U^{(i)}$ are defined as:
 \be
 \label{U_def}
 U^{(i)}_{l'm';lm}=\left(
 {\calM^{(i)}+i\over\calM^{(i)}-i}
 \right)_{l'm';lm}\;.
 \ee
 This is the general relation we are looking for
 in the case of two-channel scattering.
 Obviously, when the off-diagonal matrix elements
 of the $S$-matrix, i.e. $S^{(l)}_{12}$ and
 $S^{(l)}_{21}$ vanish,
 Eq.~(\ref{eq:main_result}) reduces to the famous
 L\"uscher's formula~\cite{luscher91:finitea}
 for the single-channel elastic scattering.

 \section{Eigenstates with definite cubic symmetry}
 \label{sec:symmetry}

 The general result~(\ref{eq:main_resultU})
 obtained in the previous section
 can be further simplified when we consider irreducible
 representations of the symmetry group of the cubic box.
 We know that energy eigenstates in a box can be
 characterized by their transformation properties under
 the symmetry group of the box.
 For this purpose, one has to decompose the representations
 of the rotational group with angular momentum $l$ into
 irreducible representations of the corresponding symmetry
 group of the box. For a symmetric cubic box,
 the relevant symmetry group is the cubic group $O(Z)$.

 In a given symmetry sector, denoted by the irreducible
 representation $\Gamma$, the representation of the
 rotational group with angular momentum $l$ is decomposed into
 irreducible representations of $O(Z)$. This decomposition
 may contain the irreducible representation $\Gamma$.
 We may pick our basis of the representation
 as: $|\Gamma,\alpha;l,n\rangle$.
 Here $\alpha$ runs from $1$ to $dim(\Gamma)$, the dimension
 of the irreducible representation $\Gamma$. Label $n$ runs from
 $1$ to the total number of occurrence of $\Gamma$ in
 the decomposition of rotational group representation with
 angular momentum $l$. The matrix $\hat{M}$ is diagonal with
 respect to $\Gamma$ and $\alpha$ by Schur's lemma.

 For the two-particle eigenstates in the symmetry
 sector $\Gamma$, the general formula~(\ref{eq:main_resultU})
 reduces to:
 \be
 \label{eq:main_resultG}
 \left|\begin{array}{cc}
 U^{(1)}(\Gamma)-
 S^{(l)}_{11}
 &
 \sqrt{{k_2m_2\over k_1m_1}}S^{(l)}_{21}
 \\
 \sqrt{{k_1m_1\over k_2m_2}}S^{(l)}_{12}
 &
 U^{(2)}(\Gamma)- S^{(l)}_{22}
 \end{array}\right|=0\;,
 \ee
 Here $\hat{U}(\Gamma)$ represents a linear operator in
 the vector space $\calH_\Lambda(\Gamma)$
 \footnote{Please refer to Ref.~\cite{luscher91:finitea} for details.}.
 This vector space is spanned by all complex vectors
 whose components are $v_{ln}$, with $l\leq\Lambda$,
 and $n$ runs from $1$ to the number of occurrence of
 $\Gamma$ in the decomposition of representation with angular
 momentum $l$~\cite{luscher91:finitea}.
 To write out more explicit formulae,
 one therefore has to consider decompositions
 of the rotational group representations under appropriate
 cubic symmetries.

 The basic symmetry group for a cubic box
 is the group $O(Z)$, which has $2$ one-dimensional
 (irreducible) representations $A_1$ and  $A_2$, a
 two-dimensional irreducible representation $E$, and
 $2$ three-dimensional representations $T_1$ and $T_2$.
 \footnote{The notations of the irreducible representations
 of group $O(Z)$ that we adopt here follow those in
 Ref.~\cite{luscher91:finitea}.}%
 Up to angular momentum $l=4$,
 the representations of the rotational group
 are decomposed according to:
 \ba
 \label{eq:decomposition_OZ}
 {\mathbf 0} &=& A^+_1\;,\nonumber \\
 {\mathbf 1} &=& T^-_1\;,\nonumber \\
 {\mathbf 2} &=& T^+_2+E^+\;, \\
 {\mathbf 3} &=& A^-_2+T^-_1+T^-_2\;,\nonumber \\
 {\mathbf 4} &=& A^+_1+E^++T^+_1+T^+_2\;,\nonumber
 \ea
 In most lattice calculations, the symmetry sector that is
 easiest to investigate is the invariant sector: $A^+_1$.
 We therefore focus on this particular symmetry sector.
 We see from Eq.~(\ref{eq:decomposition_OZ}) that,
 up to $l\leq 4$, only $s$-wave and $g$-wave contribute to this sector.
 This corresponds to {\em two} linearly independent, homogeneous polynomials
 with degrees not more than $4$ which are invariant under $O(Z)$.
 The two basis polynomials can be identified as
 $\calY_{00}$ and $\calY_{40}$.
 \footnote{Our conventions for the spherical harmonics are
 taken from Ref.~\cite{jackson:book}.}

 In the first order approximation,
 if we neglect the mixing between
 the $s$-wave and $g$-wave, we have for the $A^+_1$ sector:
 \be
 \label{eq:swave_only}
 \det\left(\begin{array}{cc}
 e^{2i\Delta_1}-\eta_0e^{2i\delta^0_1} &
 i\sqrt{{k_2m_2\over k_1m_1}}
 \sqrt{1-\eta^2_0}e^{i\delta^0_1+i\delta^0_2} \\
 i\sqrt{{k_1m_1\over k_2m_2}}
 \sqrt{1-\eta^2_0}e^{i\delta^0_1+i\delta^0_2} &
 e^{2i\Delta_2}-\eta_0e^{2i\delta^0_2}
 \end{array}\right)=0\;.
 \ee
 where we have also used the special
 parametrization~(\ref{eq:S_parametrize})
 for the $s$-wave $S$-matrix elements and
 we have defined:
 \be
 e^{2i\Delta_i}\equiv {\calM_{00}(k^2_i)+i
 \over \calM_{00}(k^2_i)-i}\;,
 \ee
 which may also be expressed as:
 \be
 \label{eq:Delta_def}
 \cot\Delta_i=\calM_{00}(k^2_i)
 ={\calZ_{00}(1,q^2_i) \over \pi^{3/2} q_i}\;.
 \ee
 Note that the quantities appearing above,
 namely $\delta^0_1$,  $\delta^0_2$, $\eta_0$,
 $\Delta_1$ and $\Delta_2$, are all functions of the energy:
 $E=k^2_1/(2m_1)=E_T+k^2_2/(2m_2)$.
 Expanding Eq.~(\ref{eq:swave_only}) we get,
 after some algebra:
 \be
 \label{eq:swave_simple}
 \cos(\Delta_1+\Delta_2-\delta^0_1-\delta^0_2)=
 \eta_0\cos(\Delta_1-\Delta_2-\delta^0_1+\delta^0_2)\;.
 \ee
 This is the simplified formula for the $s$-wave
 $S$-matrix elements. Another equivalent way of writing
 the same formula is:
 \be
  \label{eq:swave_simple2}
 \tan(\Delta_1-\delta^0_1)\tan(\Delta_2-\delta^0_2)
 ={1-\eta_0\over 1+\eta_0}\;.
 \ee
 Therefore, if we neglect contaminations
 from higher angular momentum (mainly from $l=4$),
 the parameters in the two-channel $S$-matrix
 elements, namely $\eta_0$, $\delta^0_1$, $\delta^0_2$
 and the exact two-particle energy $E$ satisfy
 a relation given by~(\ref{eq:swave_simple}).
 Unlike the single channel case, where the
 $S$-matrix has only one parameter (phase shift) and
 it is related to the exact energy in a one-to-one fashion,
 the two channel $S$-matrix now has $3$ real parameters and
 these parameters are related to the exact energy $E$
 by one relation. This relation is helpful since
 it provides a constraint on the four physical quantities.
 For example, if we can measure the exact energy $E$ in
 lattice calculations, and if we know the values
 of $\delta^0_1$ and $\delta^0_2$
 from experimental data (e.g. by partial wave analysis),
 we in principle can infer information about
 the parameter $\eta_0$ which is
 difficult to measure in the experiment.
 If the experimental information is inadequate, say both
 $\delta^0_2$ and $\eta_0$ are poorly determined,
 our result~(\ref{eq:swave_simple}) still helps to
 setup a constraint between the two poorly determined
 physical quantities. This is more or less the
 situation in $\pi\pi$ scattering just
 above the $KK$ threshold.
 Note that above the two-particle inelastic threshold,
 the number of states within a particular energy interval
 is roughly twice as many as in the
 single-channel case and every energy eigenvalue satisfies
 Eq.~(\ref{eq:swave_simple}).

 It is instructive to discuss the situation just above the
 inelastic threshold. Assuming that the inelastic scattering
 only occurs in the $s$-wave,
 the total reaction cross section just above the
 threshold is given by:
 \be
 \sigma^{l=0}_r \simeq A\sqrt{E-E_T}\;.
 \ee
 for very small $(E-E_T)>0$ where $A$ is some
 proportionality constant. This means that the parameter $\eta_0$
 behaves like:
 \be
 \eta_0\simeq 1-{Am_1E_T\over \pi}\sqrt{E-E_T}\;,
 \ee
 just above the threshold. On the other hand, since
 the physical quantity $\delta^0_2$ vanishes at the
 threshold, we may parameterize it as:
 \be
 \tan \delta^0_2\simeq k_2a^{(2)}_0\;.
 \ee
 Quantity $a^{(2)}_0$ might be called the
 scattering length in the second channel.
 Substituting these into our general
 formula~(\ref{eq:swave_simple2}), we get:
 \footnote{Note that $\Delta_2\sim q_2/\calZ_{00}(1,q^2_2)\sim k^3_2$ for
 small $k_2$, so $\tan(\Delta_2-\delta^0_2)\sim -\tan\delta^0_2$.}
 \be
 a^{(2)}_0={m_1E_T A\over 2\pi\sqrt{2m_2}}
 \cot\left[\delta^0_1(E_T)-\Delta_1(E_T)\right]
 \;.
 \ee

 If we consider the mixing of the $g$-wave, the formula
 obtained above becomes more complicated.
 We can write out the four-dimensional reduced matrix
 $\calM(A^+_1)$ whose matrix elements are denoted as:
 $\calM(A^+_1)_{ll'}=m_{ll'}=m_{l'l}$, with $l$ and $l'$ takes values
 in $0$ and  $4$, respectively.
 Using the general formula~(\ref{eq:calM-zeta}), it is
 straightforward to work out these reduced matrix elements
 in terms of matrix elements $\calM_{lm;l'm'}$.
 We find that, in the case of cubic symmetry,
 Eq.~(\ref{eq:main_resultG}) becomes:
 \be
 \label{eq:4by4}
 \left|\begin{array}{cccc}
 u^{(1)}_{00}-S^{(0)}_{11} &
 \sqrt{{k_2m_2\over k_1m_1}}S^{(0)}_{21} &
 u^{(1)}_{04} &  0\\
 \sqrt{{k_1m_1\over k_2m_2}}S^{(0)}_{12} &
 u^{(2)}_{00}-S^{(0)}_{22} &
 0 & u^{(2)}_{04} \\
 u^{(1)}_{04} &  0 &
 u^{(1)}_{44}-S^{(4)}_{11} &
 \sqrt{{k_2m_2\over k_1m_1}}S^{(4)}_{21}\\
 0 & u^{(2)}_{04} &
 \sqrt{{k_1m_1\over k_2m_2}}S^{(4)}_{12} &
 u^{(2)}_{44}-S^{(4)}_{22}
 \end{array}\right|=0
 \;.
 \ee
 Here matrix $U$ is defined as in Eq.~(\ref{U_def}) and
 $u_{ll'}$ are the corresponding matrix elements.
 At this level, so many parameters enter the relation and
 it seems that the formula is useful
 only when other information is available.

 We also would like to remark that, although we worked out
 the formulae in a cubic box, similar relations can also
 be obtained for general rectangular box following
 the strategies outlined in
 Ref.~\cite{chuan04:asymmetric,chuan04:asymmetric_long}.
 Such topologies might be useful for the calculations
 of scattering phases. It is also clear that the
 results obtained in this paper can easily generalized to
 more than two channels as long as the initial and final
 states in the scattering are still two-particle states.

 Finally, let us speculate about possible extension to
 the case of massive field theory. In the case of single
 channel scattering, it was shown in Ref.~\cite{luscher91:finitea}
 that the result obtained in the quantum-mechanical model
 can be carried over literally to the case of massive
 quantum field theory as long as the
 non-relativistic dispersion relations of the particles
 are replaced by relativistic ones and the
 polarization effects and other effects (exponentially small)
 are small enough. For the case of multi-channel scattering,
 we expect that a similar conclusion to hold although a
 proof is still lacking. If this turned out to be true,
 it means that the results obtained in this paper can
 also be generalized to the case of massive field theory
 apart from corrections that are exponentially small in
 the large volume limit.

 In the case of field theory, another complication arises since
 in the framework of field theory, particle numbers are not conserved.
 One therefore has to specify what one means
 by two-particle states~\cite{luscher91:finitea} in a finite volume.
 Generally speaking, two-particle states for elastic scattering
 are the discrete (but quasi-continuum) spectrum
 states above the lowest two-particle
 threshold. When the energy is increased, we may
 encounter another threshold where the inelastic
 channel is opened. In the case of pion-pion scattering,
 the lowest two-particle threshold is the two-pion threshold.
 The next lowest threshold is the four-pion threshold which
 is below the two-kaon threshold. It is clear that the formulae
 obtained in this paper would be applicable only when the coupling
 between the four pion states and the two particle (two pion and two kaon)
 states becomes negligible. If this were the case,
 then when we calculate the correlation
 function matrix among appropriate two-particle operators,
 two-particle states dominate the correlation function matrix.
 The energy eigenstates thus obtained are also mainly composed of
 two-particle states. As we have said in the introduction,
 the weakness of interaction of four pion states with two pion
 states can be seen from the chiral lagrangian in which they
 are coupled via derivative couplings.
 In a lattice QCD simulation, the validity
 of this assumption might also be checked numerically
 by investigating the volume dependence of the energy
 eigenstates obtained from the corresponding correlation functions.
 This is due to the fact that the density of state for
 single particle, two-particle  and four-particle states
 have rather different volume dependence.
 Using this technique, for example, the
 authors in Ref.~\cite{kfliu04:pentaquark} were
 able to argue that the energy
 eigenstates they obtained are in fact $KN$ scattering states
 (two-particle states) and not a single-particle penta-quark state.
 Similar technique can in principle be applied to distinguish
 the two-particle states from the four-particle states
 in a finite volume. Of course the feasibility of this
 can only be checked in a real numerical simulation. Here
 we can only point out this possibility.

 \section{Conclusions}
 \label{sec:conclude}

 In this paper, we have studied two-particle two-channel
 scattering states in a cubic box with periodic boundary conditions.
 Assuming that energy eigenstates are only two-particle states,
 the relation of the exact energy eigenvalues in the box and the
 physical parameters in the coupled channel $S$-matrix elements
 in the continuum is found. This formula
 can be viewed as a generalization of
 the well-known L\"uscher's formula to the
 coupled channel situation (inelastic scattering).
 In particular, we show that the two-channel $S$-matrix elements
 in the $s$-wave are related to the energy of the two-particle system
 by a simple identity, if contaminations from higher
 angular momentum sectors are neglected.
 This relation is non-perturbative in nature and it
 will help us to establish connections between the
 $S$-matrix parameters in the multi-channel
 scattering with the energy eigenvalues which are
 in principle accessible in lattice calculations.

\section*{Acknowledgments}

 We would like to thank Prof. H.~Q.~Zheng of Peking University
 for drawing our attention to the possibility of studying coupled
 channel scattering using lattice techniques. We also benefited greatly
 from various discussions with him.

 \appendix
 \section{Appendix A}
 \label{app:proof}

 In this appendix, the proof of the theorem concerning the structure of
 the solution to the radial Schr\"odinger equation is provided.
 In particular, we consider the coupled differential equations:
 \be
 \label{eq:radial_z}
 {d^2\Psi(z)\over dz^2}+{2\over z}{d\Psi(z)\over dz}+Q(z)\Psi(z)=0\;,
 \ee
 where $\Psi(z)$ is the two-component wave-function and
 the matrix $Q(z)$ is given by:
 \be
 Q(z)=-{l(l+1)\over z^2}+
 \left(\begin{array}{cc}
 2m_1[E-V_1(z)] & 2m_1\Delta(z) \\
 2m_2\Delta^*(z) & 2m_2[E'-V_2(z)]
 \end{array}\right)\;.
 \ee
 We would like to study the structure of the solutions
 to Eq.~(\ref{eq:radial_z}) near $z=0$.

 Using standard transformation:
 \be
 \Psi(z)=z^{-1/2}\Phi(z)\;,
 \ee
 we find that Eq.~(\ref{eq:radial_z}) reduces to:
 \be
 \label{eq:radial_Phi_z}
 {d^2\Phi(z)\over dz^2}+{1\over z}{d\Phi(z)\over dz}
 +\tilde{Q}(z)\Phi(z)=0\;,
 \ee
 with $\tilde{Q}(z)$ given by:
 \be
 \tilde{Q}(z)=-{(l+1/2)^2\over z^2}
 + \left(\begin{array}{cc}
 2m_1[E-V_1(z)] & 2m_1\Delta(z) \\
 2m_2\Delta^*(z) & 2m_2[E'-V_2(z)]
 \end{array}\right)\;.
 \ee
 We now proceed to find canonical solutions of the type:
 \be
 \label{eq:series_solution}
 \Phi(z)=z^\rho\sum^\infty_{n=0}\chi_nz^n\;,
 \ee
 where $\chi_n$ are two-component coefficients,
 $\rho$ is the index of the canonical solution.
 Assuming $z^2V(z)$ is analytic near $z=0$
 and $\lim_{z\rightarrow 0}[z^2V(z)]=0$, we have:
 \be
 z^2\tilde{Q}(z) = \sum^\infty_{n=0}Q_n z^n
 \;,\;\;
 Q_0 =-(l+1/2)^2 \;
 \ee
 Comparing the coefficients of each order we get:
 \be
 \label{eq:recursion}
 [(\rho+n)^2)-(l+1/2)^2]\chi_n
 +\sum^n_{k=1}Q_k\chi_{n-k}=0\;.
 \ee
 In particular, when $n=0$ the above equation gives
 the index equation satisfied by $\rho$:
 \be
 \rho^2-(l+1/2)^2=0\;.
 \ee
 This equation yields two solutions: $\rho=\pm(l+1/2)$.
 Note that $\chi_0$ is non-vanishing, this shows that
 the solutions near $z=0$ are categorized into two
 classes. In one class, the solutions behave like:
 $\Phi(z)\sim z^{l+1/2}$; in the other class the
 solutions behave like: $\Phi(z)\sim z^{-(l+1/2)}$.
 Or in terms of the solution $\Psi(z)$, these two
 cases behave like $\Psi(z)\sim z^l$ and
 $\Psi(z)\sim z^{-(l+1)}$ respectively.
 Also note that when the index $\rho$ take the
 values of $\pm(l+1/2)$, the matrix $Q_0$ becomes
 identically zero. Obviously, we can find two
 linearly independent solutions $\chi_0$ for
 each $\rho$. Therefore, we have shown that the
 canonical solutions indeed have the properties as
 indicated in the theorem.
 Since the coefficients $\chi_n$ are fully determined by
 the recursion relation~(\ref{eq:recursion}),
 all what remains to be
 shown is that the series solution~(\ref{eq:series_solution})
 converges uniformly and absolutely
 in the neighborhood of $z=0$.

 To show the convergence of the series solution,
 we see from Eq.~(\ref{eq:recursion}) that:
 \be
 \chi_n=\mp{1\over n(2l+1)}\sum^n_{k=1}
 Q_k\chi_{n-k}\;.
 \ee
 Note that $z^2\tilde{Q}(z)$ is
 analytic near $z=0$, therefore we can always find
 two real numbers $M\ge 1$ and $R>0$ such that:
 \be
 || Q_k || \le MR^{-k}\;,
 \ee
 where $||\cdot||$ indicates a matrix norm.
 \footnote{The concrete form of the matrix norm
 is irrelevant as long as it is a well-defined
 matrix norm that is consistent with the
 ordinary Euclidean vector norm.}
 Let us assume that:
 \be
 |\chi_\nu|\le M^\nu R^{-\nu} |\chi_0|,\;\;
 \nu=1,2,\cdots,(n-1)\;.
 \ee
 we now show that the above inequality then holds
 for $\nu=n$. This is seen by:
 \ba
 |\chi_n| &\le& {1\over n(2l+1)}\sum^n_{k=1}
 ||Q_k|||\chi_{n-k}|
 \nonumber \\
 &\le& {|\chi_0|R^{-n}\over n(2l+1)}\sum^n_{k=1}
 M^{n-k+1}
 \le M^nR^{-n}|\chi_0|\;.
 \ea
 Therefore, by induction,
 the inequality $|\chi_n|\le M^nR^{-n}|\chi_0|$ is
 true for any integer $n$.
 It then follows trivially that the
 series~(\ref{eq:series_solution}) is absolutely
 and uniformly convergent in a small neighborhood
 around $z=0$ which completes our proof of the theorem.

 %\bibliography{mybib}
 %\bibliographystyle{unsrt_nt}

 \end{document}